\documentclass[aps,showkeys,superscriptaddress,twocolumn,nofootinbib]{revtex4-1}
\usepackage{subscript}
\usepackage{natbib}
\usepackage{amssymb}
\usepackage{color}
\usepackage{amsfonts}
\usepackage{textcomp}
\newcommand{\ped}[1]{\ensuremath{_{\rm #1}}}
\newcommand{\apex}[1]{\ensuremath{^{\rm #1}}}
\definecolor{blue}{rgb}{0,0,0}
\usepackage{float}
\usepackage[caption = false]{subfig}
\usepackage{graphicx}

\begin{document}

\title{Possible charge-density-wave signatures in the anomalous resistivity of Li-intercalated multilayer MoS\ped{2}}

\author{Erik Piatti}
\affiliation{Department of Applied Science and Technology, Politecnico di Torino, corso Duca degli Abruzzi 24, 10129 TO Torino, Italy}

\author{Qihong Chen}
\affiliation{Device Physics of Complex Materials, Zernike Institute for Advanced Materials, Nijenborgh 4, 9747 AG Groningen, The Netherlands}

\author{Mauro Tortello}
\affiliation{Department of Applied Science and Technology, Politecnico di Torino, corso Duca degli Abruzzi 24, 10129 TO Torino, Italy}

\author{Jianting Ye}
\email{j.ye@rug.nl}
\affiliation{Device Physics of Complex Materials, Zernike Institute for Advanced Materials, Nijenborgh 4, 9747 AG Groningen, The Netherlands}

\author{Renato S. Gonnelli}
\email{renato.gonnelli@polito.it}
\affiliation{Department of Applied Science and Technology, Politecnico di Torino, corso Duca degli Abruzzi 24, 10129 TO Torino, Italy}

\begin{abstract}
We fabricate ion-gated field-effect transistors (iFET) on mechanically exfoliated multilayer MoS\ped{2}. We encapsulate the flake by Al\ped{2}O\ped{3}, leaving the device channel exposed at the edges only. A stable Li\apex{+} intercalation in the MoS\ped{2} lattice is induced by gating the samples with a Li-based polymeric electrolyte above $\sim 330$ K and the doping state is fixed by quenching the device to $\sim 300$ K. This intercalation process induces the emergence of anomalies in the temperature dependence of the sheet resistance and its first derivative, which are typically associated with structural/electronic/magnetic phase transitions. We suggest that these anomalies in the resistivity of MoS\ped{2} can be naturally interpreted as the signature of a transition to a charge-density-wave phase induced by lithiation, in accordance with recent theoretical calculations.
\end{abstract}


\maketitle

\section{Introduction}\label{sec:introduction}

Interacting electrons in transition metal dichalcogenides (TMDs) have attracted a lot of attention, owing to the emergence of exotic electronic phases and the non-trivial physics that arise due to their competition \cite{EfrosPollakBook}. Experimentally, these competing electronic phases are often the cause of anomalies in the temperature dependence of the electric transport - a characteristic feature of a wide variety of  materials including oxides \cite{CampbellPRB1997}, arsenides \cite{BeleanuPRB2013}, iron pnictides \cite{BaekPRB2009,WuSciRep2014}, and metal chalcogenides \cite{KatayamaSSC1976,KlemmReview2015}. These anomalies mark the boundaries between different electronic phases, exhibiting transitions in the lattice, magnetic, electronic and topological degrees of freedom. Different transitions can also appear concurrently across the same boundary in a material's phase diagram. This can be incidental, if the two transitions are  unrelated to one another \cite{CampbellPRB1997}. Alternatively, their concomitant occurrence may result from a strong coupling between the underlying phases, when the transition in one degree of freedom triggers a transition in a second one \cite{BaekPRB2009,WuSciRep2014}.

TMDs are layered compounds sharing the generalized MX\ped{2} formula, where M is a transition metal (such as Mo, Nb, Ta, Ti, V, W) and X is a chalcogen (S, Se, Te) element {\color{blue}\cite{ManzeliReview2017, ChoiReview2017}. Different structural phases are possible in these materials depending on the coordination of the metal atom and the stacking of the individual layers, the most common being trigonal prismatic (2\textit{H}), octahedral (1\textit{T}) and distorted octahedral (1\textit{T}') \cite{ManzeliReview2017, ChoiReview2017}. For most - but not all - TMDs at room temperature, the 2\textit{H} polytype is the most stable, while the other metastable polytypes can be obtained via alkali ion intercalation \cite{ManzeliReview2017, EnyashinCTC2012}. TMDs} feature complex electronic structures and complicated phase diagrams reminiscent of those of cuprates and iron pnictides \cite{KlemmReview2015}, often dominated by the interplay between superconductivity (SC) and charge-density-wave (CDW) order {\color{blue}\cite{ManzeliReview2017}}. CDWs are periodic modulations of the charge carrier density, associated to distortions of the underlying crystal lattice \cite{GrunerBook}, which are often the result of strong electron-phonon coupling \cite{RossnagelJPCM2011} and Fermi-surface nesting \cite{ZhuPNAS2015}. {\color{blue}Their appearance and behavior in TMDs are strongly dependent on both the atomic components and the polytype \cite{ManzeliReview2017}: }although well known in several Nb-, Ta-, Ti-, and V-based TMDs \cite{KlemmReview2015}, as well as metallic 1\textit{T}'-MoTe\ped{2} \cite{KeumNatPhys2015}, this phenomenon has not been reported so far in the semiconducting phase of TMDs of the 2\textit{H} polytype. Interestingly, however, the application of an external pressure in excess of $\sim10$ GPa has been recently shown to induce a metallization of 2\textit{H}-MoS\ped{2}, accompanied by the possible emergence of a CDW distortion \cite{Caoarxiv2018}.

\begin{figure*}
\begin{center}
\includegraphics[keepaspectratio, width=0.8\textwidth]{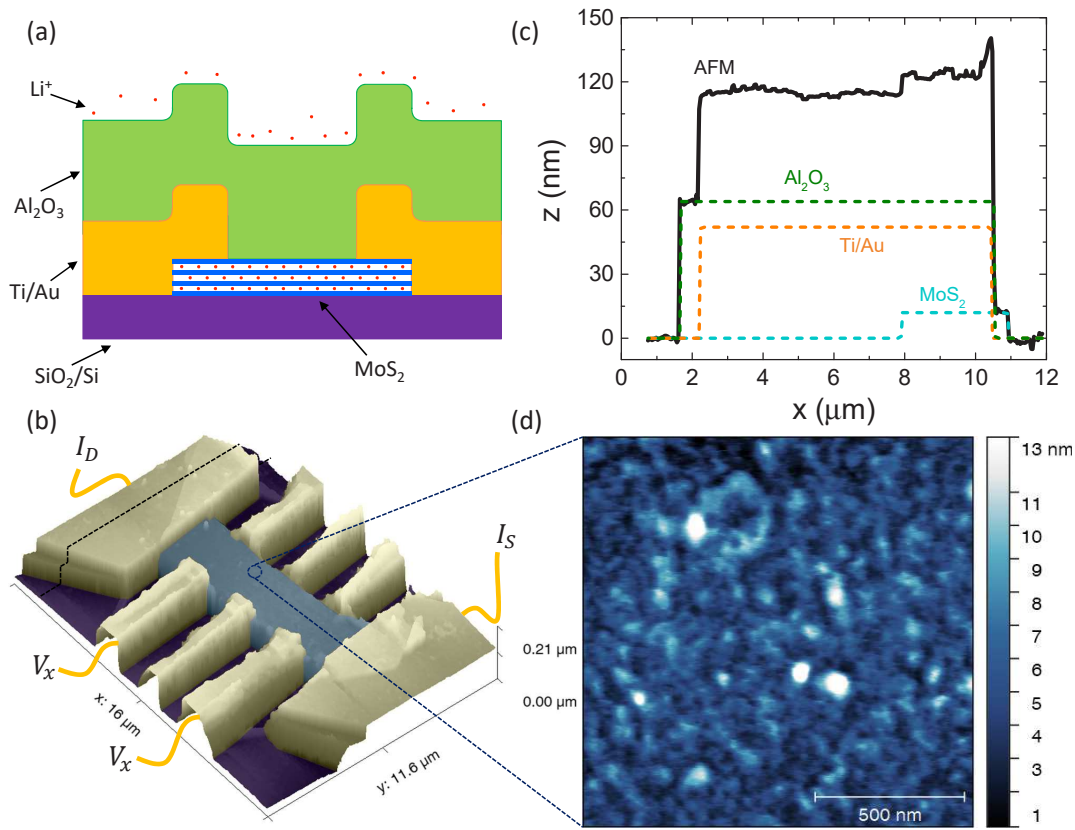}
\end{center}
\caption {
(a) Sketch in side-view and (b) 3D rendering of the AFM topographic image in tapping mode of an Al\ped{2}O\ped{3}-encapsulated few-layer MoS\ped{2} device. Li\apex{+} ions can directly access the MoS\ped{2} flake only from the exposed sides of the device. $V_G$ is applied between $I_S$ and a coplanar side-gate electrode (not shown in the figure). (c) Height signal in correspondence to the black dashed line in (b) (black solid line). Dashed lines highlight the topographic contributions from the MoS\ped{2} flake (blue), Ti/Au contacts (yellow) and Al\ped{2}O\ped{3} encapsulation layer (green). (d) AFM topographic image in tapping mode of the same device in the area highlighted by the dashed blue circle in (b). Root mean square height is $S_q=1.86$ nm, much smaller than the Al\ped{2}O\ped{3} thickness ($\sim60$ nm).} \label{figure:device}
\end{figure*}

At ambient pressure, the interplay between SC and CDWs in TMDs used to be accessed by carrier doping via  intercalation \cite{KlemmReview2015, FangPRB2005, MorosanNatPhys2006, WagnerPRB2009, BhoiSciRep2016, YuNatNano2015}. Recently, it was demonstrated that the same effect can be induced by the application of an electric field \cite{YoshidaSciRep2015, LiNature2016, XiPRL2016}. Structural transitions between different polytypes can also be controlled by electric field effect, as it has been shown in single-layer MoTe\ped{2} between the 2\textit{H} and 1\textit{T}' phases - although the authors did not investigate whether this did also give rise to CDW order \cite{WangNature2016}. Indeed, carrier doping has been predicted to be able to induce different CDW phases in MoS\ped{2} {\color{blue}\cite{RosnerPRB2014, ZhuangPRB2017, ChenCSB2013}. On the other hand, electrochemical intercalation holds great promise for tailoring its electric, optical and chemical properties, as well as making it very attractive in the field of energy storage for TMD-based alkali-ion batteries and supercapacitors \cite{ChoiReview2017}. In this sense, ionic gating can be an especially versatile tool for doping control. This technique incorporates an electrochemical cell in a top-gate transistor configuration, where the sample is separated from a gate counter-electrode by an electrolyte \cite{UenoReview2014}. By properly selecting gating temperature and gate voltage, doping can be achieved either via electrostatic ion accumulation or bulk electrochemical intercalation. In the electrostatic regime, the gate voltage drives the ions to form an electric double layer at the surface of the sample, which acts as a nanoscale capacitor with a huge capacitance \cite{LiNature2016, XiPRL2016, ShiSciRep2015}. In the electrochemical regime, the strong interface electric field is exploited to drive the ions in the van der Waals gap between the layers, achieving gate-controlled intercalation \cite{YuNatNano2015, ShiSciRep2015, PiattiAPL2017}.}

In this work, we use ionic gating with a polymeric electrolyte to intercalate Li\apex{+} ions in Al\ped{2}O\ped{3}-encapsulated multilayer MoS\ped{2} devices at high temperature. We find that the Li\apex{+}-intercalated state is fully stable upon removal of the applied gate voltage, if the sample is quenched below the optimal intercalation temperature. No doping-induced SC state is observed down to $\sim 3$ K, but we find clear evidence for the emergence of sheet resistance ($R_s$) anomalies in the intercalated state around $\sim 200$ K. These anomalies evolve as a hump-dip structure in the first derivative, $dR_s/dT$. {\color{blue}Upon increasing the gate voltage beyond the onset of intercalation, the hump feature strongly shifts to higher temperature, while the position of the dip feature remains mostly unaffected.} We discuss how this behavior can be naturally linked with the appearance of the CDW phases predicted in {\color{blue}Refs.\cite{RosnerPRB2014, ZhuangPRB2017, ChenCSB2013}} as a function of Li\apex{+} doping. To the best of our knowledge, our results constitute the first report for non-Fermi liquid behavior in MoS\ped{2} at ambient pressure.

\section{Results}

\subsection{Device fabrication}

We first developed a procedure to fabricate encapsulated MoS\ped{2} devices, where the only direct interface between the flakes and the electrolyte occurs at the sides of the flake (see Fig.\ref{figure:device}a). We obtained multilayer MoS\ped{2} flakes by mechanical exfoliation \cite{NovoselovPNAS2005} of 2\textit{H}-MoS\ped{2} bulk crystals (SPI Supplies) on standard SiO\ped{2}(300 nm)/Si substrates. Flakes with thicknesses around $\sim10$ nm (number of layers $\sim15$) were selected through their optical contrast  \cite{LiACSNano2013}, which were confirmed subsequently by atomic force microscopy (AFM). We chose this particular thickness to study the well-defined bulk properties \cite{SplendianiNanoLett2010}, while minimizing the effect of lattice expansion in the \textit{z} direction during the Li\apex{+} intercalation, which can easily break the electrodes if the expansion is too severe in thicker samples. Electrical contacts to the flakes were patterned in Hall bar configuration by e-beam lithography, followed by evaporating Ti(5 nm)/Au(50 nm) and lift-off. A large, interdigitated coplanar side-gate electrode was patterned $\sim100\,\mu$m away from the flake \cite{PiattiAPL2017}. Then, we deposited a Al\ped{2}O\ped{3}($\sim60$ nm) mask over the electrodes and the rectangular channel of the Hall bar, leaving the irregular part of the flake exposed on all sides. Finally, we employed reactive ion etching (using Ar gas, RF Power: 100 W, etching duration: 2 min) to remove the exposed areas of the flake.

Fig.\ref{figure:device}b shows a 3D rendering of the AFM height signal acquired in tapping mode over a completed device before applying the electrolyte. We use false colors to clearly distinguish the different regions of the device (yellowish gray: leads, blue: channel, and violet: substrate). The Al\ped{2}O\ped{3} encapsulation layer covers both the channel and the leads. Along the leads, it partially extends on the underlying substrate to provide complete insulation from the environment. On the device channel, it presents the sharp edge defined by RIE allowing direct exposure of the flake to the electrolyte only from the side. The stacking of the three materials, as sketched in Fig.\ref{figure:device}a, can be clearly recognized from the height profile of the AFM imaging (Fig.\ref{figure:device}c) along the black dashed line in Fig.\ref{figure:device}b. Height steps corresponding to the MoS\ped{2} flake, the Ti/Au contacts and the Al\ped{2}O\ped{3} encapsulation mask can be clearly recognized and are explicitly highlighted.

In Fig.\ref{figure:device}d, we present the surface topography of the channel region. The smooth Al\ped{2}O\ped{3} encapsulation layer on atomically flat MoS\ped{2} is free of pinholes or other defects that would allow penetration of ions from the  electrolyte to the channel surface. Direct AFM profiling for a typical $1.5\times1.5\,\mu$m$^2$ area shows that the root mean square roughness of the Al\ped{2}O\ped{3} surface $S_q$ equals 1.86 nm, which is more than 30 times smaller than the total oxide thickness.

For the intercalation experiments, we prepared the polymeric electrolyte by dissolving $\sim25$ wt.\% of lithium bis(trifluoromethane)sulfonimide (Li-TFSI) in polyethylene glycol (PEG, $M_w\sim450$) in an Ar-filled glove box. For the low-temperature control experiment (described in the following), we directly employed the ionic liquid 1-butyl-1-methylpiperidinium bis(trifluoromethylsulfonyl)imide (BMPPD-TFSI). Both electrolytes are liquid at room temperature, and the latter retains good ionic mobility down to $\sim240$ K. The electrolytes are pumped under vacuum at $\sim330$ K for at least 1 hour before being drop casted on to the device, covering both the channel and the Au side-gate electrode. Subsequently, the devices are quickly transferred to the cold plate of a Cryomech\textregistered PT405 pulse-tube cryocooler and allowed to degas under high vacuum ($\lesssim 10^{-5}$ mbar) for another 1 hour to minimize water absorption. This is necessary to eliminate the electrolysis of absorbed water, which might be activated easily before the expected intercalation due to the combination of high temperature and large applied gate voltage.

\subsection{Electrochemical doping}

\begin{figure*}
\begin{center}
\includegraphics[keepaspectratio, width=0.8\textwidth]{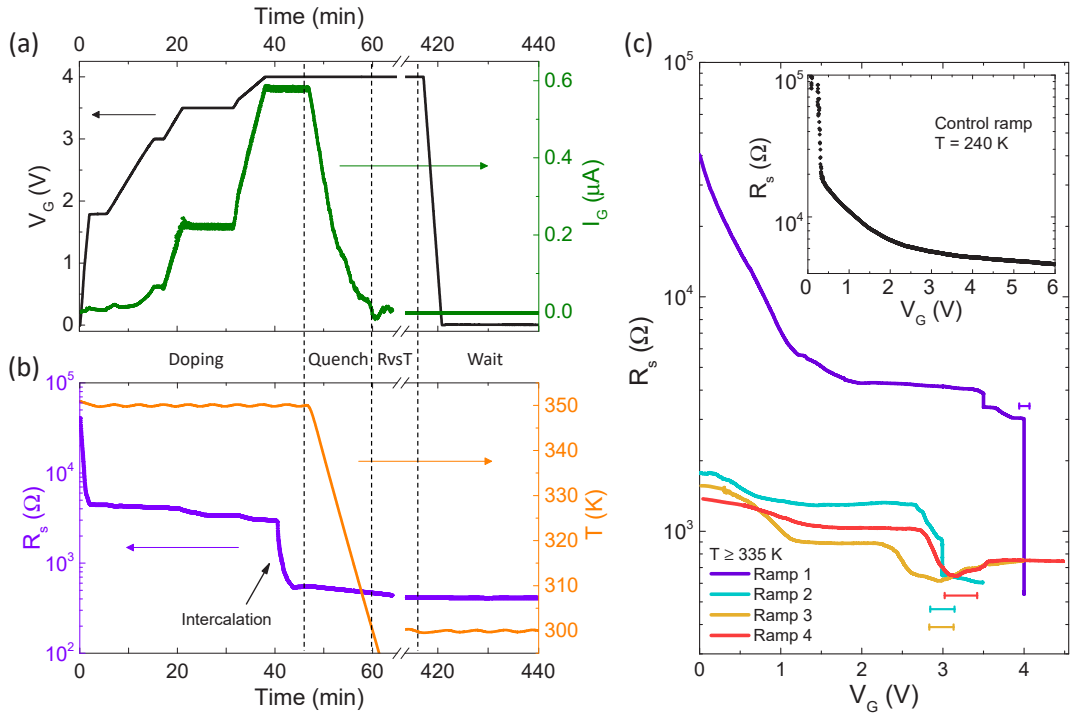}
\end{center}
\caption {
Encapsulated MoS\ped{2} doping dynamics. (a) $V_G$ (black line, left axis) and $I_G$ (green line, right axis) as a function of time during the doping process, quenching, characterization, and release. (b) $R_s$ (violet line, left axis) and $T$ (orange line, right axis) as a function of time. Dashed lines distinguish the different experimental steps. (c) $R_s$ as a function of $V_G$ for different $V_G$ ramps. Ramp 1 is measured at $T\simeq350$ K. Ramps 2,3,4 at $T\simeq335$ K. All are performed using the PEG/Li-TFSI polymer electrolyte. Horizontal error bars indicate the uncertainty on the threshold voltage $V_{th}$ for stable Li\apex{+} intercalation in each ramp. Inset shows a $V_G$ ramp measured at $T\simeq240$ K using pure BMPPD-TFSI ionic liquid for comparison.} \label{figure:doping}
\end{figure*}

We perform Li\apex{+} intercalation in our encapsulated MoS\ped{2} devices by applying gate voltage $V_G$ ramps at high temperature ($T\gtrsim 330$ K) and monitoring the sheet resistance $R_s$ as a function of time. We employ the first channel of a two-channel Agilent B2912 Source-Measure Unit (SMU) to apply $V_G$ and measuring the gate current $I_G$ simultaneously. Then the $R_s$ is determined in the four-wire configuration by applying a small constant DC current between the source and drain contacts of our devices ($I_{DS}\sim1\mu$A) with the second channel of the same SMU, and measuring the longitudinal voltage drop $V_{xx}$ across two voltage contacts with an Agilent 34420 Nanovoltmeter. We remove the common mode errors, such as thermoelectric offsets and contributions from $I_G$, by averaging the $R_s$ values acquired with the $I_{DS}$ of opposite polarities (two-point delta mode).

As shown in Fig.\ref{figure:doping} a and b, the overall gating process can be divided into four main steps: doping, quenching, $T$-dependent characterization, and release. In the first step, the $V_G$ is ramped following the profile shown in Fig. 2a at high $T$ = 330 K and then kept constant for typically $\sim15$ min allowing the insertion of ions into the flake. In the second step, the doping process in quenched by cooling the sample below the optimal intercalation temperatures while keeping the $V_G$ constant. In the third step, the full temperature dependence of the $R_s$ is investigated. This will be discussed in detail in the next section. In the fourth and final step, the $V_G$ is ramped down to zero: depending on the $T$, at which this step is performed, the intercalated ions can be released back to the electrolyte, or remain confined in the MoS\ped{2} lattice.

As shown in Figs.\ref{figure:doping}b and \ref{figure:doping}c, for the modulation of the $R_s$ as a function of $V_G$, the gating process can be separated into two regimes. For the low $V_G$, the modulation of $R_s$ is mainly originated from driving the Li\apex{+} ions by the applied electric field and accumulating them electrostatically onto the channel surface \cite{ShiSciRep2015,PiattiAPL2017}. For large values of $V_G$, the change in $R_s$ is instead mainly caused by field-driven ion intercalation to the van der Waals gap between the MoS\ped{2} layers \cite{ShiSciRep2015,PiattiAPL2017}. When the MoS\ped{2} top surface is directly exposed to the electrolyte, it is difficult to separate the two regimes by considering only the behavior of $R_s$ \cite{PiattiAPL2017}, therefore, additional information can be useful for clearer discrimination. Most notably the carrier density in the sample determined by Hall effect \cite{PiattiAPL2017} can be used as an useful guidance. On the other hand, when the top surface of the flake is protected by the Al\ped{2}O\ped{3} layer, the encapsulation strongly suppresses the electrostatic gating on the device channel. Hence, the two regimes are clearly separated by the sharp $R_s$ drop appearing when the ions penetrate between the layers \cite{YuNatNano2015}.

Subsequently, the doping process is quenched by rapidly cooling down the sample. This procedure ``locks'' the intercalated ions in place and leads to a stable lithiation state for the MoS\ped{2} flake. Indeed, once $V_G$ is released to zero at a lower $T\sim300$ K, no significant change in $R_s$ can be detected on a time scale between tens of minutes to a few days. Note that, at this $T$, the electrolyte is still liquid and hence fully supports ion motion: thus, this behavior is qualitatively different from the ``freezing'' of the ions electrostatically accumulated in the form of the electric double layer when the system is cooled below the glass transition temperature of the electrolyte. The $R_s$ increases again over time at $V_G = 0$ only when the sample is heated above $T\gtrsim320$ K, signaling the onset of delithiation. Hence, lithiation-delithiation of the MoS\ped{2} flake can be achieved above $320$ K: after trial-and-error, we found out that gating at $T\sim335$ K provided efficient doping while minimizing the chance of device failure at high $V_G$.

In Fig.\ref{figure:doping}c we show different $V_G$ ramps, corresponding to different intercalation states. These are achieved by selecting  different target $V_G$ values while keeping the same doping time. However, the $R_s$ drop clearly shows that, between successive ramps, the onset of intercalation does not remain unchanged, and is instead affected by the previous intercalation history of the flake. The different intercalation states can then more properly be mapped by how much the final applied $V_G$ \emph{exceeds} the onset of intercalation in that specific ramp, \textit{i}.\textit{e}. the ``overdrive" voltage $V_G - V_{th}$, where $V_{th}$ is the value of $V_G$ where a stable Li\apex{+} incorporation is achieved. For ramps where $R_s$ drops while $V_G$ is not held constant, we choose $V_{th}$ as the value of the minimum in the $R_s$ vs. $V_G$ curve. Hence, the electrostatic regime is identified by the condition $V_G - V_{th}<0$, while the intercalation one by $V_G-V_{th}\gtrsim0$.

\begin{figure*}
\begin{center}
\includegraphics[keepaspectratio, width=0.8\textwidth]{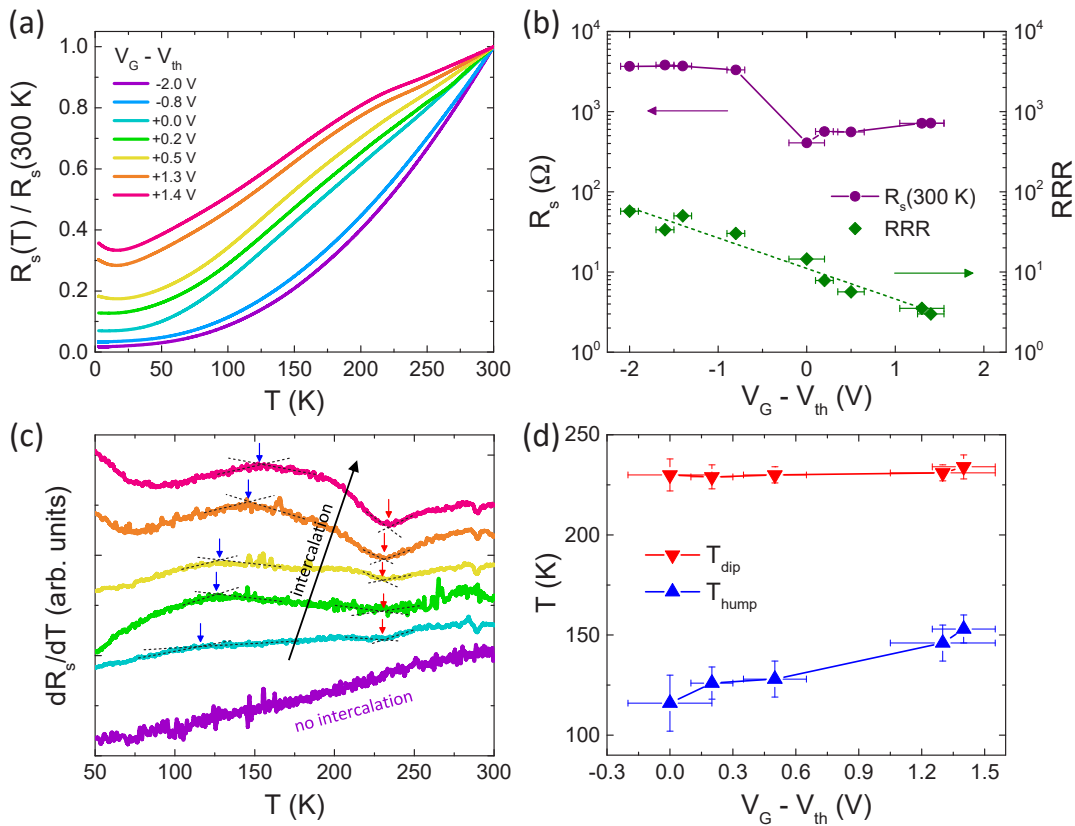}
\end{center}
\caption {
Temperature dependence of the electrical transport. (a) $R_s$, normalized at $T=300$ K, as a function of $T$, before and after the onset of Li\apex{+} intercalation. (b) Overdrive voltage dependence of the $R_s$ at $T=300$ K (violet dots and lines), and of the residual resistivity ratio (green diamonds). Dashed line acts as a guide to the eye. (c) $T$ dependence of the first derivative of the $R_s$, before and after Li\apex{+} intercalation. {\color{blue}Curves are color-coded to match the legend in panel (a).} Arrows indicate the $T$ values where the first derivative shows a dip (red) and a peak (blue). Curves are shifted for clarity. (d) Overdrive voltage dependence of $T_{hump}$ (blue up triangles) and $T_{dip}$ (down red triangles) in (c).} \label{figure:transport}
\end{figure*}

Finally, we confirm that the $R_s$ modulation at low $V_G$ is due to electrostatic accumulation. To show this, we performed a control doping experiment using an ionic liquid and applied $V_G$ at $T\sim240$ K. It is well known that reducing the gating temperature strongly suppresses all electrochemical interactions maintaining pure electrostatic charging \cite{UenoReview2014, ShiSciRep2015, YeScience2012, PiattiarXiv2018}. Furthermore, due to much larger ions in the molecular ionic liquids, gate-driven intercalation of layered flakes by ionic liquids is known to result in immediate device failure due to delamination and destruction of the crystal structure \cite{YuNatNano2015}. As we show in the inset to Fig.\ref{figure:doping}c, gating an encapsulated device with pure BMPPD-TFSI ionic liquid results in a featureless, monotonically decreasing dependence of $R_s$ on $V_G$, and a comparable $R_s$ modulation with respect to the $V_G-V_{th}<0$ regime of Li-TFSI gating at $T\gtrsim330$K. In both cases the flakes quickly revert to their native insulating states by simply removing $V_G$, further supporting the electrostatic picture. Note that an Al\ped{2}O\ped{3} layer with a thickness $d\sim60$ nm ($\epsilon_r\sim9$ \cite{BiercukAPL2003}) provides a residual, non-negligible top-gate capacitance $C_{ox}=\epsilon_r\epsilon_0/d\sim130$ nF/cm\apex{2}, where $\epsilon_0$ is the vacuum permittivity. It is worth noting that an additional electrostatic contribution may also arise from quasi-1D channels induced at the exposed sides of the MoS\ped{2} flakes.

We now show the electric transport of our encapsulated devices down to $T\sim3$ K, both before ($V_G-V_{th}<0$) and after ($V_G-V_{th}\gtrsim0$), the onset of intercalation. Fig.\ref{figure:transport}a shows the $T$-dependence of the resistance, normalized by its value at $300$ K. The samples show clear metallic behavior both in the electrostatic and intercalated regimes. In Fig.\ref{figure:transport}b we plot two figures of merit for the transport properties as a function of $V_G-V_{th}$: the sheet resistance at $300$ K, $R_s(300\mathrm{K})$, and the residual resistivity ratio $RRR$ defined as $R_s(300 \mathrm{K})/R_{s,0}$, where $R_{s,0}$ is the minimum of $R_s$ over the entire measured $T$ range. The behavior of $R_s(300 \mathrm{K})$ well reproduces our observation that, during the doping process, encapsulating the flake with Al\ped{2}O\ped{3} is crucial to clearly distinguish surface and bulk-doped states as shown by the sharp drop in $R_s$ close to $V_{th}$. The decreasing dependence of $RRR$ instead indicates that carrier mobility is strongly suppressed by increasing $V_G$, especially in the intercalation regime. This behavior is possibly affected by the randomness caused by intercalation and progressive introduction of extra scattering centers due to the presence of the ions themselves, irrespectively of the material under study \cite{GallagherNatCommun2015, OvchinnikovNatCommun2016, PiattiApSuSc2017, Gonnelli2DMater2017, PiattiAPL2017}.

For $V_G-V_{th}<0$, $R_s$ is a smooth increasing function of $T$. The large $RRR$ values in this regime indicate that defects provide a small contribution to the total carrier scattering rate at high $T$. This can be associated to the insulation of the device from the ionic environment provided by the encapsulation layer, and is consistent with the drastically enhanced carrier mobility reported in MoS\ped{2} transistors encapsulated with Al\ped{2}O\ped{3} \cite{KimAPL2016} and other high-$\kappa$ dielectrics \cite{KuferNanoLett2015}. For $V_G-V_{th}\geq0$, however, a clear change of slope appears in the temperature dependence of $R_s$ around $\sim 200$ K. This anomalous ``hump'' becomes more evident for larger values of $V_G-V_{th}$. Below this hump ($T\lesssim 200$ K), the $R_s$ drops more rapidly with the decrease of $T$. At the same time, we observe the emergence of a resistance upturn for $T\lesssim 20$ K. This indicates that, at low $T$ and high doping, metallic behavior is suppressed.

The behavior of the resistance hump around $\sim 200$ K can be best visualized by the $T$ dependence of the first derivative of $R_s$, $dR_s/dT$, both before and after intercalation as shown in Fig.\ref{figure:transport}c. When the sample is not intercalated {\color{blue}(violet curve)}, $dR_s/dT$ is a featureless function of $T$. In the intercalated state {\color{blue}(upon increasing doping: light blue, green, yellow, orange and red curves)}, the resistance anomaly gives rise to a clear dip-hump structure in the $T$ dependence of $dR_s/dT$: a sharp dip at higher $T$ ($T_{dip}$, highlighted by the red arrows), and a broader hump at lower $T$ ($T_{hump}$, blue arrows). These two features evolve differently with increasing doping: as we show in Fig.\ref{figure:transport}d, the $T_{hump}$ strongly increases with increasing $V_G-V_{th}$, while the $T_{dip}$ is nearly constant.

\section{Discussion}

Resistance anomalies in TMDs are usually associated with phase transitions to various CDW phases \cite{KlemmReview2015}. These are ubiquitous in Nb-, Ta- and Ti-based dichalcogenides for both main polytypes (trigonal prismatic 2\textit{H} and octahedral 1\textit{T}) \cite{KlemmReview2015}, with the exception of NbS\ped{2} \cite{NaitoJPSJ1982}. In particular, CDW transitions are observed in the $T$ dependence of the resistivity of undoped TMDs as large, hysteretic jumps in insulating compounds \cite{WilsonAdvPhys1975,SiposNatMater2008,YoshidaSciRep2015}, and as less apparent humps in metallic ones \cite{NaitoJPSJ1982,TidmanPhilMag1974,XiNatNano2015}.

Experimentally, doping can control CDW phases in TMDs, both by field-effect carrier accumulation at the surface \cite{YoshidaSciRep2015,LiNature2016,XiPRL2016} as well as ion intercalation in the bulk \cite{YuNatNano2015,MorosanNatPhys2006,BhoiSciRep2016,FangPRB2005,ZhuJPCM2009}. Generally, increase of carrier doping causes strong suppression of the CDW phases, favoring the onset of SC order \cite{YoshidaSciRep2015, LiNature2016, YuNatNano2015, MorosanNatPhys2006, BhoiSciRep2016, FangPRB2005, ZhuJPCM2009}. This, however, is not true for all compounds: for example, it has recently been demonstrated that doping strengthens the CDW phase both in 2\textit{H}-NbSe\ped{2} and 2\textit{H}-TaSe\ped{2} \hbox{thin flakes \cite{XiPRL2016}}.

We thus consider whether the resistance anomalies we observed in encapsulated Li\ped{x}MoS\ped{2} could be attributed to the emergence of a CDW phase. Such an interpretation would naturally account for the two different features observed in $dR_s/dT$ at $T_{hump}$ and $T_{dip}$, as well as their doping dependence. Indeed, in other TMDs, the dip in $dR_s/dT$ is associated to a transition to an incommensurate CDW phase at higher $T$, which is weakly dependent on doping \cite{BhoiSciRep2016}, while the peak in $dR_s/dT$ to the further transition to a commensurate CDW phase at lower $T$, with a strong doping dependence \cite{BhoiSciRep2016}. In addition, Ref.\cite{RosnerPRB2014} predicted that sufficiently strong electron doping ($\gtrsim 0.15$ e\apex{-}/cell) would also cause the suppression of the SC dome in 2\textit{H}-MoS\ped{2}, and the appearance of CDW order due to phonon instabilities. Conversely, Ref.\cite{ZhuangPRB2017} calculated that both electron and hole doping may trigger a structural phase transition in MoS\ped{2}, from the semiconducting 2\textit{H} phase to the metallic 1\textit{T} phase. Furthermore, they predicted that while hole doping stabilizes the metastable 1\textit{T} phase, electron doping would then promote the transition to the more stable, distorted 1\textit{T}' phase. The latter can be regarded as a CDW restructuring of the metallic 1T phase and should exhibit semimetallic behavior with a graphene-like Dirac cone in absence of spin-orbit coupling \cite{ZhuangPRB2017}.

{\color{blue}Strictly speaking, these theoretical results were calculated for a single-layer. Nevertheless, we expect multilayer samples to follow a qualitatively similar behavior, since bulk lithiation has been explicitly predicted to promote CDW transitions in both the 2\textit{H} and 1\textit{T} polytypes \cite{ChenCSB2013}. Notably, CDW order is expected to be weaker in the 2\textit{H} structure, with moderate lattice distortion and suppressed electron localization, while the opening of a full band gap is predicted for the 1\textit{T} structure \cite{ChenCSB2013}.} Indeed, the TMD phase engineering by lithium intercalation has explicitly been demonstrated \cite{VoiryCSR2015}: in the case of MoS\ped{2}, lithiation of the pristine 2\textit{H} structure can result in the formation of both the regular 1\textit{T} and distorted 1\textit{T}' structures, as observed in STEM and Raman measurements \cite{EdaNanoLett2012,LengACSNano2016}. A similar behavior was observed in Re-doped MoS\ped{2} as well \cite{LinNatNano2014}.

A further consistency check between the behavior of our samples and the onset of a CDW phase can be obtained by assessing the scaling of the $T$ dependence of the $R_s$ in the intercalated state. In 2\textit{H}-NbSe\ped{2}, TaS\ped{2} and TaSe\ped{2}, the CDW ordering is phenomenologically associated with a pronounced change in the slope of the $T$ dependence of $R(T)$, when plotted in log-log scale. At high $T$, the scaling is linear in $T$ due to large-angle carrier scattering by acoustic phonons \cite{NaitoJPSJ1982}. At low $T$, the scattering follows a $T^p$ dependence due to small-angle electron-phonon scattering instead \cite{NaitoJPSJ1982}. The value of $p$ depends on the orbital symmetry of the bands involved in the scattering process: scattering from \textit{s}-like bands, such as in 2\textit{H}-TaS\ped{2} and TaSe\ped{2}, gives $p=5$ \cite{NaitoJPSJ1982}, while scattering from \textit{d}-like bands, such as in 2\textit{H}-NbSe\ped{2}, gives $p=3$ \cite{NaitoJPSJ1982}. Larger values of $p$ are also considered as a measure of stronger CDW strength \cite{NaitoJPSJ1982}. For the $T$ between these extremes, scaling with an intermediate $p\simeq2$ is expected corresponding to the scattering by CDW fluctuations \cite{NaitoJPSJ1982}.

\begin{figure}
\begin{center}
\includegraphics[keepaspectratio, width=0.9\columnwidth]{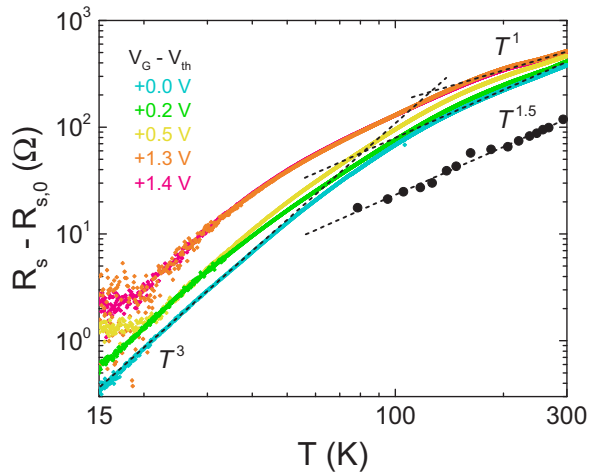}
\end{center}
\caption{
Double logarithmic plot of the increase of $R_s$ in the range $15-300$ K for Li\apex{+}-intercalated MoS\ped{2} (small diamonds) and K\ped{0.4}MoS\ped{2} (black dots) from Ref.\citenum{HermannSSC1973}. Dashed lines that represent the $T^1$, $T^{1.5}$, and $T^3$ power-law dependences are shown for comparison.} \label{figure:scaling}
\end{figure}

As shown in Fig.\ref{figure:scaling}, for $T\gtrsim 200$ K, our devices show a $T^p$ scaling with $p$ between $1$ and $1.5$. This reproduces well the behavior in bulk intercalated samples described in literature \cite{HermannSSC1973} (black dots in Fig.\ref{figure:scaling}). The slightly super-linear scaling can be originated from both doping inhomogeneity \cite{HermannSSC1973} and additional scattering with optical phonons \cite{YuAdvMater2017}. For $T\lesssim 40$ K, and neglecting the localization upturn close to the lowest $T$, the curves present a scaling very close to $p=3$. This is consistent with the dominant orbital \textit{d}-character of the conduction band of MoS\ped{2} \cite{LiuPRB2014}. Intermediate values of $T$ show a further scaling behavior with $p\simeq$ 1.6 to 2.1.

As substantiated by the discussion above, our results are consistent with a doping-induced CDW ordering in Li\ped{x}MoS\ped{2}. However, transport measurements alone cannot distinguish between different possible scenarios leading to such a behavior. In the simplest scenario, that is proposed in Ref.\cite{RosnerPRB2014}, Li\apex{+} ions simply provide charge carriers to the system, which undergoes a CDW transition by a reconstruction of the 2\textit{H} structure (similarly to what was observed under external pressure \cite{Caoarxiv2018}). This would place the CDW phase in competition with SC order present at lower doping. We point out that no sample that presented the $R_s$ anomalies showed any kind of SC transition. This is in contrast to our earlier report \cite{PiattiAPL2017}, where Li\ped{x}MoS\ped{2} flakes showed SC at $T\lesssim 3.7$ K while not showing $R_s$ anomalies at higher $T$. There are two main differences between these experiments. On one hand, in the present work the flakes are encapsulated, which eliminates the possibility of inducing superconductivity by field effect. Indeed, encapsulation has been shown to promote CDW ordering in TMDs \cite{YanAPL2015}. More importantly, the devices in Ref.\cite{PiattiAPL2017} were intercalated at a lower $T=300$ K, thus strongly suppressing ionic mobility in the flake. Thus, the devices presented in this work may all show larger doping levels even at $V_G-V_{th}\simeq0$, placing their state well beyond the peak of SC dome, and pushing $T_c$ well below the lowest $T$ accessible in our experiment (3 K).

Another possibility is that the larger intercalation $T\gtrsim 330$ K employed in this work allowed for a structural phase transition away from the pristine 2\textit{H} polytype {\color{blue}\cite{ZhuangPRB2017, ChenCSB2013}}. Thus, a second scenario would more closely follow the picture proposed in Ref.\cite{ZhuangPRB2017}. Doping with Li\apex{+} ions would first induce a structural transition of the MoS\ped{2} from the 2\textit{H} to the 1\textit{T} phase. Then the metastable 1\textit{T} phase would undergo a CDW transition into the distorted 1\textit{T}' phase at lower temperatures. Alternatively, doping could also induce a transition directly from the 2\textit{H} to the 1\textit{T}' phase{\color{blue}, as theoretically suggested in Ref.\cite{EnyashinCTC2012}}. In this case, the observed transport anomalies would require the presence of multiple CDW distortions available for the system, which have been predicted for 1\textit{T}'-MoTe\ped{2} \cite{LeeArXiv2017}. Finally, we cannot rule out the possibility that the sample remains in the 2\textit{H} phase at high $T$ and undergoes a purely structural transition to the 1\textit{T}/1\textit{T}' phases during the cool down. However, we deem this last possibility unlikely due to the metastable nature of the 1\textit{T} phase and the energy barrier that separates the two polytypes even in presence of Li\apex{+} intercalants \cite{ZhuangPRB2017}, which would hinder a transition at lower temperatures. These points cannot be settled purely by electric transport measurements, and further work will require using structure-sensitive  techniques to characterize Li\ped{x}MoS\ped{2} both as a function of $T$ and doping, such as Raman spectroscopy or X-ray diffraction studies.

\section{Conclusions}
In summary, we fabricated multilayer MoS\ped{2} devices to investigate the effects of field-driven Li\apex{+} intercalation with a polymeric electrolyte at $T\gtrsim330$ K. To minimize the influence of ion accumulation at the surface, we encapsulated our devices with Al\ped{2}O\ped{3} high-$\kappa$ dielectric, and employed RIE to obtain well-defined device geometry and edges, leaving only the sides of the flake exposed to the electrolyte. The resulting device architecture was confirmed via AFM. We monitored the effects of field-driven Li\apex{+} intercalation by measuring $R_s$ in our devices as a function of $V_G$ and $T$. Encapsulation allowed us to clearly distinguish between surface ion accumulation and bulk ion intercalation by the presence of sharp drops in the $R_s$ vs. $V_G$ and time curves. We confirmed the electrostatic operation of our devices before the onset of intercalation by gating with a pure ionic liquid at $T\sim 240$ K. The doping process at high $T$ resulted in stable Li\apex{+} incorporation in the MoS\ped{2} lattice even when the electrolyte was still liquid, as long as the sample was then quenched below $320$ K. We characterized the $T$ dependence of $R_s$ in the intercalated samples down to $\sim 3$ K, and observed anomalous metallic transport with a doping-induced hump in $R_s$ around $T\sim 200$ K. These anomalous features strongly suggest the onset of a possible phase transition in the intercalated flakes, and are the first report of anomalous metallic character in MoS\ped{2} at ambient pressure. In analogy with the behavior of several other TMDs, and in accordance with theoretical predictions from the literature, we propose an interpretation of these anomalies in terms of the formation of a CDW phase at large Li\apex{+} doping.

\section*{Acknowledgments}
Q. Chen and J. T. Ye acknowledge funding from the European Research Council (Consolidator Grant No. 648855 Ig-QPD).

\end{document}